\documentclass[longbibliography,aps,reprint,onecolumn,10pt,showpacs,showkeys,preprintnumbers,amsmath,amssymb,floatfix,superscriptaddress]
{revtex4}

\usepackage{graphicx} 
\usepackage{dcolumn}  
\usepackage{bm}       
\usepackage{hyperref}
\usepackage{color}
\usepackage{epsfig}
\usepackage{float}

\newcommand{\ul}{\underline}

\newcommand{\beq}{\begin{equation}}
\newcommand{\eeq}{\end{equation}}
\newcommand{\ds}{\displaystyle}

\begin{document}

\title{\normalsize Volume in memory of Prof. Mauro Francaviglia \\ 
\Large 
Gauge dependence and self-force from Galilean to Einsteinian free fall, compact stars falling into black holes, Hawking radiation and the Pisa tower at the general relativity centennial
}

\author{Alessandro D.A.M. Spallicci\footnote{Chaire Fran\c{c}aise, Universidade do Estado do Rio de Janeiro, 
Instituto de F\'isica, Departamento de F\'isica Te{\'o}rica, Rua S\~{a}o Francisco Xavier 524, Maracan\~{a}, 20550-900 Rio de Janeiro, 
Brasil}}
\email{spallicci@cnrs-orleans.fr}
\homepage{http://lpc2e.cnrs-orleans.fr/~spallicci/}
\affiliation{Universit\'e d'Orl\'eans \\ 
Observatoire des Sciences de l'Univers en r\'egion Centre\\
P\^ole de Physique, Collegium Sciences et Techniques\\ 
Centre Nationale de la Recherche Scientifique\\
Laboratoire de Physique et Chimie de l'Environnement et de l'Espace, UMR 7328\\
3A Avenue de la Recherche Scientifique, 45071 Orl\'eans, France}

\author{Maurice H.P.M. van Putten}
\affiliation{Sejong University\\
Department of Astronomy and Space Science\\
209 Neungdong-ro Gwangjin-gu, 143-747 Seoul, Korea}

\date{9 July 2016}
\begin{abstract}
Obviously, in Galilean physics, the universality of free fall implies an inertial frame, which in turns implies that the mass $m$ of the falling body is omitted (because it is a test mass; put otherwise, the centre of mass of the system coincides with the centre of the main, and fixed, mass $M$; or else, we consider only an homogeneous gravitational field). Otherwise, an additional (in the same or opposite direction) acceleration proportional to $m/M$ would rise either for an observer at the centre of mass of the system, or for an observer at a fixed distance from the centre of mass of $M$.
These elementary, but overlooked, considerations fully respect the equivalence principle and the (local) identity of an inertial or a gravitational pull for an observer in the Einstein cabin. They value as fore-runners of the self-force and gauge dependency in general relativity. Because of its importance in teaching and in the history of physics, coupled to the introductory role to Einstein's equivalence principle, the approximate nature of Galilei's law of free fall is explored herein. When stepping into general relativity, we report how the geodesic free fall into a black hole was the subject of an intense debate again centred on coordinate choice. Later, we describe how the infalling mass and the emitted gravitational radiation affect the free fall motion of a body. The general relativistic self-force might be dealt with to perfectly fit into a geodesic conception of motion. Then, embracing quantum mechanics, real black holes are not classical static objects any longer. Free fall has to handle the Hawking radiation, and leads us to new perspectives on the varying mass of the evaporating black hole and on the varying energy of the falling mass. Along the paper, we also estimate our findings for ordinary masses being dropped from a Galilean or Einsteinian Pisa-like tower with respect to the current state of the art drawn from precise  measurements in ground and space laboratories, and to the constraints posed by quantum measurements. The appendix describes how education physics and high impact factor journals discuss the free fall. Finally, case studies conducted on undergraduate students and teachers are reviewed. 
\end{abstract}


\maketitle
\vspace{-0.65 cm}
\hspace{1.55 cm}{\footnotesize Mathematics Subject Classification 2010: 83.01} 

\section{Introduction and motivations}

Radial fall was a source of inspiration instantiated by the stone of Aristot\'el\=es, the tower of Galilei, the apple of Newton, and the cabin of Einstein. The equivalence principle (EP) affirms  that inertial mass and gravitational weight are proportional;  such proportionality - fixed to unity - is independent of the kind of matter. The universality of free fall (UFF) implies that equal weights, regardless of their composition, fall with equal  accelerations. 
The EP and the UFF  don't affirm though, that, not just test bodies, but lighter and heavier weights of the same composition fall at the same rate, regardless of the reference frame. It would be an incorrect statement both in pre-relativistic and relativistic physics.   
 
Whether the EP is really a principle, it is a doubt to many theorists, and Einstein himself initially considered the EP only as hypothesis 'Aequivalenzhypothese'. But as long as it holds, the UFF is a testable consequence. We don't discuss the microscopic matter composition and the different contributions of the fundamental interactions here, {\it e.g.}, the Nordtvedt effect \cite{no72} is not relevant herein \footnote{{The literature on the EP is immense. For a technical introduction oriented to experimental verification see \cite{will93,will14}; for a famous classical criticism on the limits on the equivalence of inertial and gravitational masses see \cite{sy60}; for a quantum modern analysis see \cite{fayet01,da12}}; for  pedagogical work, for the relations among the various EP definitions see Di Casola et al. \cite{dcliso15}, while Ohanian \cite{oh77,oh79} and Walstad \cite{walstad1979} clarify the concept of test bodies and the locality of the EP. }. Instead, we remind that bodies of same material, but different mass, don't fall equally, if not approximately. Test bodies (or put differently, when equating the centre of mass of the system to that of the main mass, being the latter fixed or else when we consider only an homogeneous gravitational field) fall with the same accelerations simply because we have neglected the influence of their weights. 
Truly, in the context of pre-relativistic mechanics, the issue of mass dependent acceleration has rarely been addressed.
In the appendix, we show that when dealing with Galilean free fall, high impact science and physics education journals, as well as media, not always state that the uniqueness of free fall is solely valid in an ideal inertial frame. This restriction is known by many, but definitely not to all students, and certainly not exploited. 

Previous physics education literature has identified an education gap between high school programmes and university courses. On top, the media generate a myth and a know-it-all posture, {\it e.g.}, through shallow experiments - the Apollo 15 worldwide display of the simultaneous fall of a feather and a hammer \cite{apollo15} or the dropping of differently sized water bottles from the Pisa tower to fest the 400\textsuperscript{th} anniversary of Galilei's telescope and the International Year of Astronomy \cite{macisaac12,shore} - and, via careless wording, contribute to feed misconceptions relentlessly. 

But much more importantly, the inclusion of $m/M$ effects and thereby an additional (in the same or opposite direction) acceleration, might be used to introduce the concept of gauge, dominating in general relativity. That physics is ruled by a relativistic conception, where observers have different results according to their frame, is not easy to accept for undergraduates. The latter might be helped, if a gauge dependent physics is presented to them in a pre-relativistic realm.   

In the second part, we indeed pass to Einsteinian gravitation. 
With the advent of general relativity, the Earth - the reference for thought experiments - was replaced by the black hole. We examine one century of general relativity through the radial infall, starting with test bodies. Their motion is dictated by the classic geodesic equation.  The coordinate dependence that somewhat surprisingly appears in Galilean physics, now takes a predominant role in general relativity. For many decades and mainly up to the '80s, scholars debated how a test body, or a photon, falls into a black hole. Again, we let emerge the role of coordinates when pointing to the so-called repulsion effect, which occurs when coordinate time is used either with proper or non-proper length. The `discovery' of repulsion has been claimed repeatedly by several authors, who ignored the work of their predecessors.

In the '70s, the emission of gravitational radiation produced by massive particles has been analysed. Recently the influence of the emitted radiation together with the mass of the falling body have been evaluated for the motion. Herein, the term self-force will be adopted. The term is used in general relativity, but it can be presented in an elementary way already in Newtonian physics. In general relativity, self-force indicates the body reaction to its own mass and the gravitational radiation emitted. In Newtonian physics, the absence of radiation renders the self-force only related to the mass of the falling body. The concept of Newtonian self-force appeared in advanced literature, {\it e.g.}, \cite{depo04,drmewhde04,de05,de08,de11,sp11,spri14,riaospco2016b}. 

Let be clear on the meaning of self. A single body infinitely far away from any other gravitating body or external influence of any sort, will not experience any interaction with its own field and mass. Self-force is the back-reaction of a body to its mass, motion (and radiation in case of general relativity) via the intermediate role of an external field.
Incidentally, self-force in relativity is not limited to a reaction to gravitational radiation. The non-radiating modes (the spherical harmonics characterised by l=0,1) are also associated to the self-force. 

We cannot avoid remarking that the mass dependency of the acceleration represents a factor of continuity between the Galilean and Einsteinian free falls. Remarkably, we can forge the EP to hold for all these situations: we can always come up with a scenario such that the observer in the Einstein cabin will be unaware if he is under a gravitational or an inertial pull, even if its own mass is accounted for. Indeed, the general relativistic self-force MiSaTaQuWa equation \cite{misata97,quwa97}, that presents the self-force as perturbing right hand-side term of the geodesic equation, can be recast by shifting the self-force to the left-hand side, in DeWh form \cite{dewh03,spriao14}. In this way, we return to the concept of geodesic motion where no acceleration occurs, even when considering the mass of the falling body. The falling body crosses a metric given by the background plus the perturbations produced by its own mass and radiation, and a local inertial frame is identified at the particle all along the world-line.        
Self-force research has advanced largely for its relevance in building waveform templates for the detection of gravitational waves emitted by compact stars or star-sized black holes captured by supermassive black holes. 
A {\it gedankenexperiment} in a modern frame is then given by a small body captured by a supermassive black hole, and this astrophysical scenario replaces epistemologically the stone falling to the ground of pre-relativistic physics.  
These systems do generate gravitational waves and indeed are targeted by future space laser interferometer missions \cite{elisa}.  

Radial infall is used also in the context of evaporating black holes \cite{harlow2014}. We report that recently \cite{vanputten2015b}, a constant of motion has been proposed in this scenario, for which the setting up of a geodesic equation has been instrumental.    
Its combination with the Hawking radiation, leads us to new perspectives on the varying mass of the evaporating black hole and on the varying energy of the falling mass. 

Along the paper, we deal with measuring the differences in the fall of unequal weights from an ordinary height. When estimating the quantities related to a Pisa-like tower experiment, we compare them with the state of the art technology for the measurement of accelerations, distance and time, both for a non-relativistic and a relativistic fall.  

\section{Fall of unequal masses of same composition in pre-relativistic physics} 

\subsection{The gauge freedom and the Newtonian self-force}

Not questioning the EP and the UFF herein, we replace `weight' by `mass', see \cite{tarush15} as recent work dealing with the interpretation of weight. 
Indeed, nor the EP nor the UFF state that two small bodies of different mass 
$m_1$ and $m_2$ accelerate equally in a field of a large mass $M$. In Newtonian mechanics, the displacement of the centre of mass (from the centre of $M$ to a new location dependent upon the small mass $m$) corresponds to a self-force taking into account the finitude of the small mass. In other coordinates, one considers also the acceleration of $M$ towards $m$. To masses in  circular motion of radius $r_c$, the Newtonian self-force induces faster or slower angular speeds than the nominal value $\omega = \sqrt{GM/r_c^3}$, $G$ being the gravitation constant. At first order, the difference amounts to $\pm n m/M$, where the sign and the value of $n$ depend upon the chosen coordinate system \cite{de11}. 

The preceding has been extended to radial fall \cite{sp11}. 
Often authors, {\it e.g.}, \cite{nolucrshtupecaanza13}, presenting the equality of inertial and gravitational mass, write (for $d$ being the distance between the two centres of mass)

\beq
\ddot {d} m_i = - \frac{GMm_g}{d^2}~~{\rm and~thus}~~ \ddot {d} = - \frac{GM}{d^2}~,  
\label{nobili13}
\eeq 

What is implied in Eq. (\ref{nobili13}) is either that the centre of the system is coincident with the centre of mass of the main body or else that the latter if fixed. Stated otherwise, we have neglected the value of the small mass. Still for $m=m_i=m_g$, we now detail two different cases, according to the position of the observer. In both cases, the initial distance $d$ between the two bodies rests the same. We suggest to complement Eq. (\ref{nobili13}) with the following considerations. In the first case, the observer is comoving with the centre of mass which is shifted according to the value of $m$; in the second, the observer is at fixed distance from the centre of $M$, with which he accelerates towards $m$.   

{\subsubsection{Centre of mass observer (CoMO)}\label{CoMO}}

We refer to an observer at rest at the origin of the (non-inertial) centre of mass frame. 
We examine two objects of equal mass $M$ falling radially to each other, and  denote the distance of the masses to be $d=2r$. The
centre of mass $C$ of this system is in the middle of the line connecting both masses, the distance from
$C$ to each mass being $r$. The non-inertial frame $S$ is connected with $C$, and we do not consider
any external forces on the system. The acceleration of each mass with respect to $S$ is $¨\ddot r = - GM/d^2 = - GM /(2r)^2$. If we substitute any of the masses by a heavier or lighter mass $m$, one might be tempted to state that its acceleration with respect to  $S$ will stay the same, and that the acceleration of a freely falling point-like mass does not depend on its mass. But the non-inertial frame $S$ will not be centred in $C$ any longer, but elsewhere, let say $C'$, according to the value of $m$. 
The distance from $C'$ to $m$ is now $r_m$, while the distance from $C'$ to $M$ is now $r_M$. In this frame, we have 
$m r_m = M r_M$. The acceleration $\ddot {r}_m$ of $m$, is given by 


\beq
\ddot {r}_m = - \frac{GM}{d^2} 
= 
- \frac{GM}{(r_m + r_M)^2} 
= 
- \frac{GM}{r_m^2(1 + m/M)^2} 
\sim  - \frac {GM (1 - 2m/M)}{r_m^2}~.
\label{slow2}
\eeq 

Thus heavier bodies fall slower! Let $m$ be the falling stone, and $M$ the Earth. An observer comoving with the centre of mass of the system (stone plus Earth) would observe $1$ kg mass falling faster than a $2$ kg mass. 

Visually, for increasing $m$ the observer at the centre of mass get closer to $m$ itself and measures a lesser acceleration.
Finally, it is a sort of repulsion that obliges heavier bodies to fall slower for this kind of observers. In Sect. (\ref{cgi}), we shall find coordinate effects in general relativity also producing repulsion. 

One straight consideration is sufficient to cancel any Newtonian self-force: simply imposing the centre of mass of the system to be coincident with that of the Earth, that is neglecting the small value of the infalling mass, {\it i.e.} the inertial frame. But this is not the only penalty to pay. In order to recover the inertial frame, we need to superpose the centre of mass of the system on the Earth centre. But the superposition along the time of fall will depend upon the value of the small mass $m$, and thereby it is non-unique, and non-universal. Uniqueness of free fall loses its meaning in this respect, while the Newtonian self-force manifests itself as centre of mass displacement.

\vskip15pt

\subsubsection{Ground observer (GO)}\label{GO}

We deal now with a non-inertial observer on the ground, and, generally speaking, with an oberver at fixed distance from the centre of 
$M$.  
In the previous case, when considering the mutual acceleration of two equal bodies, we get $(\ddot r + \ddot r) = \ddot  d = - 2GM/d^2$. Again we replace one mass $M$ with a smaller body $m$, keeping fixed the initial separation $d$; thus $\ddot d = - G(M+m)/d^2$, and also here, we can let emerge the concept of self-force. For $h$ being the altitude of the initial position of $m$ and $r_\oplus$ the Earth radius, we find the reference system that supports a popular  stand, that is the heavier, the faster the fall will be. Indeed, for $d = r_m + r_M = h + r_\oplus$, one gets, standing on the ground, {\it e.g.}, at the feet of a Pisa-like tower ($r_\oplus$ is time independent) 

\beq
{\ddot h} = - \frac{GM (1 + m/M)}{(h + r_\oplus)^2} \sim - \frac{GM (1 + m/M)}{r_\oplus^2}~. 
\label{fast}
\eeq

Equation (\ref{fast}), might be interpreted as mutual acceleration of the large and small masses. Nevertheless, it confirms that on the ground, the observer will perceive bodies falling proportionally to their mass, in this coordinate system. This time the Newtonian self-force manifests itself as mutual acceleration, for which heavier objects get to ground sooner. 

\onecolumngrid
\begin{center}
\begin{figure}[h!]
    \centering
    \includegraphics[width=0.6\linewidth]{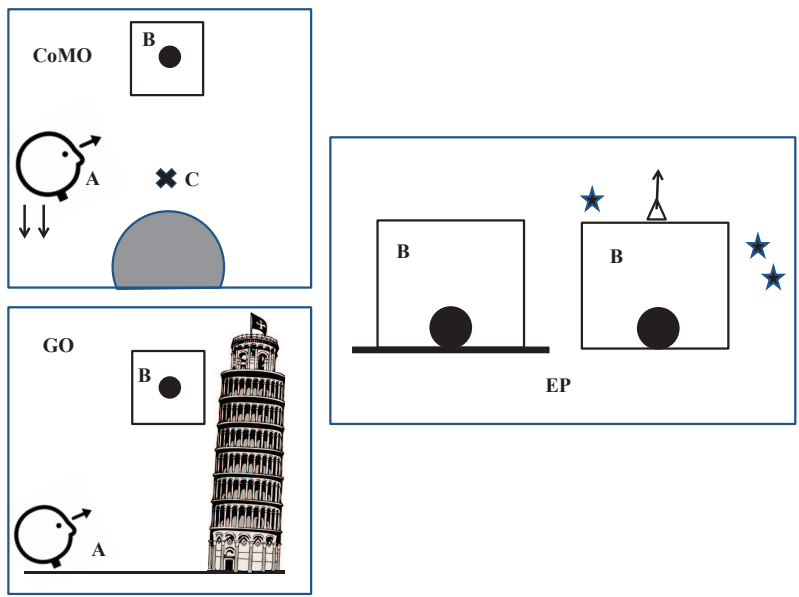}
    \caption{Lower panel - left hand-side - box GO, an observer A, fixed at the ground, sees a body, inside a one-way mirror cabin, falling with an acceleration given by Eq. (\ref{fast}). 
Upper panel - left hand-side - box CoMO, an observer A, comoving with the center of mass C of the system, sees a body, inside a one-way mirror cabin, falling with an acceleration given by Eq. (\ref{slow2}). 
In both cases, the observer B, inside the cabin, is freely floating. Freed from all forces, he does not accelerate, until he hits the ground. At that instant, right hand-side - box EP, B feels suddenly pulled downward. B, on the opaque side of the one-way mirror cabin, cannot make a difference between a gravitational downward pull and an inertial upward pull, applied at the roof cabin, supposedly in outer space, if the two pulls were  identical. He is stuck at the floor of the cabin anyway.  
There is no difference with the usual presentation of the EP, even if the small mass enters in the description. B could be subjected to a gravitational acceleration that is mass dependent in a given coordinate system; and notably, we can always arrange to pull up the cabin with the same mass dependent intensity such that B is unaware of what is going on. The self-force in general relativity includes the small falling mass in the geodesic motion by adopting the same conceptual lines, while keeping compliance to the EP.}
    \label{Fig1}
\end{figure}
\end{center}

\begin{center}
\begin{figure}[h!]
    \centering
    \includegraphics[width=0.6\linewidth]{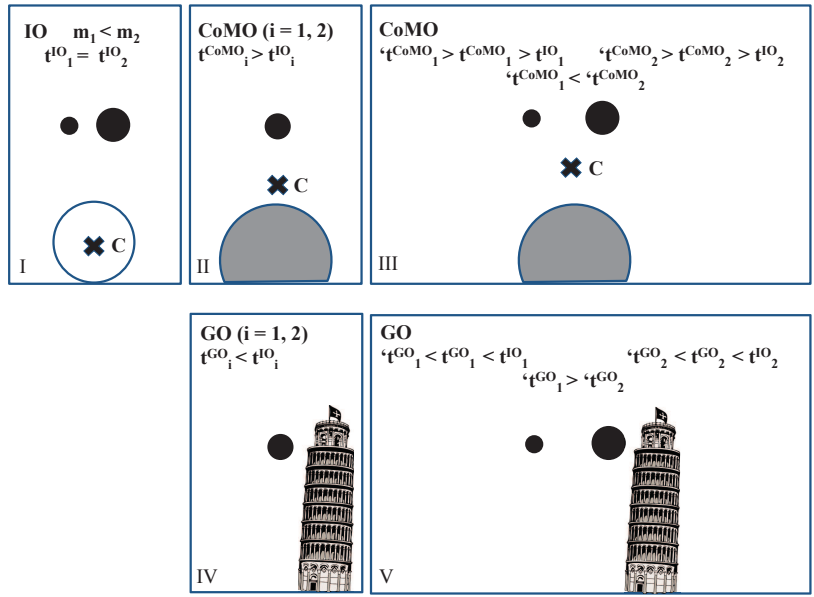}
    \caption{ 
Panel I. Inertial observer (IO). 
The centre of mass of $M$ is imposed to be coincident with the centre of mass of the system. The mass $M$  is fixed and not accelerating. Consequently, all masses have the same falling time, because we have neglected any their influence. 
Panel II. Centre of mass observer (CoMO). 
The centre of mass of the system is not any longer coincident with that of $M$, and shifts with the observer upwards proportionally to the value of the mass $m$. Accordingly, the observer is under the impression to see a lower acceleration of $m$, Eq. (\ref{slow2}). A mass $m_i$, where $i = 1, 2$, has a larger falling time than in the IO frame. Masses are always released at the same distance from the centre of mass of $M$, and one by one. 
Panel III. Centre of mass observer (CoMO). The prime indicates the simultaneous presence of both small bodies.
Masses are always released at the same distance, as in Panel II, from the centre of mass of $M$, but simultaneously. In the presence of both bodies, the centre of mass of the system shifts upwards proportionally to the mass $m_1 + m_2$. Accordingly, the observer sees lower accelerations by $m_1$ and $m_2$. The falling times will be higher than if the masses would have been released one by one. Further, the falling times will differ according to the  factor $2m_1/M$ or $2m_2/M$. 
Panel IV. Ground observer (GO). The mass $M$ is not any longer fixed and accelerates with the observer upwards proportionally to the value of the mass $m$. Accordingly, the observer is under the impression to see a higher acceleration of $m$, Eq. (\ref{fast}). A mass $m_i$ where $i = 1, 2$ has a smaller falling time than in the IO frame. Masses are always released at the same altitude from the surface of $M$, and one by one. 
Panel V. Ground observer (GO). The prime indicates the simultaneous presence of both small bodies.
Masses are always released at the same altitude, as in Panel IV, from the surface of $M$, but simultaneously. In the presence of both bodies, the mass $M$  accelerates proportionally to the mass $m_1 + m_2$. Accordingly, the observer sees higher accelerations by $m_1$ and $m_2$. The falling times will be lower than if the masses would have been released one by one. Further, the falling times will differ according to the  factor $- m_1/M$ or $-m_2/M$.
The preceding is compatible with the EP, as explained in the text and in Fig. (\ref{Fig1}).}
\label{Fig2}
\end{figure}
\end{center}

\subsection{Discussion}

It is delicate to communicate the concept of Newtonian self-force and even more, the dependence of the value and sign on the coordinates, Eqs. (\ref{slow2},\ref{fast}): a sort of pre-relativistic gauge freedom. Ironically, this has been baptised the `confusion gauge' in the annual Capra meeting on radiation reaction \cite{sp11}.    

On Earth, the Newtonian self-force is negligible ($\approx 1.7 \times 10^{-25}$~m s$^{-2}$ for each kg). Unless otherwise stated, we don't 
consider air friction and variation of acceleration with height.
For a fall - in vacuum - of $3.45$ seconds (the altitude of the Pisa tower being 58.36 meters), we should have clocks capable to appreciate fractions of yoctoseconds ($3 \times 10^{-26}$ s). Best clocks are nowadays targeting $10^{-18}$ s, therefore, eight orders of magnitude away from testing the difference between the falls of $1$ kg and $2$ kg masses. But atomic clocks have made great strides since their start in 1955, improving in accuracy by a factor of $10$ or so each decade. Will the myth on Galilei's free fall vanish in a century from now? 
Maybe even earlier, if eight orders of magnitude and eighty years will be enough. 

What about distance measurement? Ground gravitational wave interferometers of some km length are capable to measure displacements in the order of $10^{-19}$ m. When the heavier body touches ground, the lighter one (always supposing a difference of $1$ kg) lags or leads (according to the observer) one yoctometer ($ 10^{-24}$ m), in a Pisa tower-like setting. Here we have five orders of magnitude to recover.  

Strictly from this viewpoint, and thus leaving aside the composition and the nature of the falling materials, there is no surprise for a null result at $ 7 \times 10^{-9}$ in $\Delta g/g$, being $g$ the Earth gravity \cite{pechch99}, and no expectation from the EP test at $10^{-15}$ accuracy in $\mu$-SCOPE \cite{tomelero12}, as well as from the comparison of the free fall of quantum objects of different composition in STE-QUEST \cite{altschuletal2015}. 

A digression on the ESA Rosetta Mission is worth. It was launched in March 2004, rendezvoused with comet 67P/C-G in August 2014. The lander Philae touched down on the comet on 12 November and came to rest after bouncing twice. 
The comet has a mass $M = 10^{13}$ kg and a radius of $2 \times 10^4$ m, while the lander has a mass $m = 10^2$ kg.  The fall occurred from a distance of $2 \times 10^4$ m.
The free fall time would have lasted $41$ hours, but an exact computation (without self-force) demands to consider the variation of the acceleration with $r$ ($8$ hours less), and the non-vanishing initial velocity ($26$ hours less). This leads to $7$ hours of fall \cite{philae}. The effect of the Newtonian self-force amounts to half a second, easily measurable by a wrist watch, but hidden by satellite and natural multiple sources of errors and uncertainties .       

Incidentally, the ratio $m/M$ might greatly change orbits in the Solar System. In the case of the Moon-Earth system, neglecting the mass of the Moon yield a period $4$ hours too long \cite{kibble1985}.   

In conclusion, for any coordinate system, where two different masses - released from the same height - fall towards the Earth, their acceleration will be different, unless we disregard their mass, {\it i.e.} the inertial frame at the centre of the main mass. But strictly speaking such inertial observers don't exist, being rather an idealisation of non-inertial observers. In other words, bodies fall with the same acceleration only for non-existing observers. 

The deficiency of the counter-arguments consists in pushing away all conditions that would invalidate the uniqueness of free fall. Of course, in case of the test bodies, the equivalence principle holds. But a test body is a particle deprived of its mass. All bodies equally fall, because we have neglected their basic feature: their mass. 

Two comments on the equivalence of inertial and gravitational pulling. The EP states that the equivalence is valid at one point, as tidal forces show up in case of extended bodies. We bypass the remark by dealing with point masses. One might object that the point masses, which we refer to herein, are also an idealisation of real extended bodies, exactly like the inertial frame is an idealisation; furthermore, there are many instances in physics where simplifying concepts are used. We are not studying, though, real bodies (density and shape of the masses) in a real environment (air resistance, varying acceleration with altitude), but instead are concerned by the consequences of the perception of an experimental approximation - Galilei's free fall - as  principle of physics.  
   
The other comment concerns the famous cabin of Einstein. How do we account the falling mass? We have three observers: the inertial one, Eq. 
(\ref{nobili13}), who is in an inertial frame, while the other two are shown in Fig. (\ref{Fig1}): an observer comoving with the centre of mass, Eq. (\ref{slow2}); and an observer at the feet (or at the top) of the tower, Eq. (\ref{fast}). In all circumstances, there is no breaking of the EP, even if the small mass enters in the description. 
In all cases, the observer inside the falling cabin is freely floating. Freed from all forces, he does not accelerate, until he hits the ground. At that instant, he feels suddenly pulled downward, and cannot make a difference between a gravitational downward pull and an inertial upward pull, applied at the roof cabin, if the two pulls were  identical. He is stuck at the floor of the cabin anyway.  
He could be subjected to a gravitational acceleration that is mass dependent in a given coordinate system; and notably, we can always arrange to pull up the cabin with the same mass dependent intensity, such that he is unaware of what is going on. 

The small mass $m$ enters perturbatively at higher order through a coupling factor dependent of the reference frame. Interestingly, we shall see a similar situation in the relativistic self-force description of motion.    

When considering simultaneous falls, Fig. (\ref{Fig2}), we examine the following cases for both reference systems, Sect.s (\ref{CoMO},\ref{GO})
\begin{itemize}
  \item { The masses $m_1$ and $m_2$ are bound. The  difference from the inertial fall will be ruled by the factor $k(m_1 + m_2)/M$, and we record a common falling time. If we consider two new bound masses $m_3$ and $m_4$, the factor will be $k(m_3 + m_4)/M$, where either $k = -2, +1$.}
  \item { The masses $m_1$ and $m_2$ are not bound, and we neglect the gravitational interaction between them. We suppose they are axially symmetric along the line of fall, and that two centres of mass coincide, like in $\mu$-SCOPE \cite{tomelero12}. 
We are confronted with a three-body problem (a syzygy) where the influence of the masses $m_1$ and $m_2$ on their own motion is to be included.\\
Let refer to the centre of mass of the three-body system. The prime indicates the simultaneous presence of both small bodies. In a CoMO frame, Eq. (\ref{slow2}), $r_{m1}$ and $r_{m2}$ decrease due to the contemporary presence of $m_1$ and $m_2$, and are labelled $r_{m1}'$ and $r_{m2}'$, while $r_{M}$, now $r_{M}'$, increases. At start $r_{m1}' = r_{m2}':= r_m'$; we remind that the sum $r_{m} + r_M = r_{m}' + r_M' $, where $i = 1,2$, is a constant in our setting. Both bodies will have a lower acceleration than in the inertial case, and also lower than if they had been alone, but their fall will soon differ as the self-force is proportional to either $-2m_1/M$ or $-2m_2/M$. This will separate the bodies and determine $r_{m1}' \neq r_{m2}'$ and prompt a three-body system. All in all, during fall the two bodies will be subjected to different pullings.  \\
Let see the matter for a GO frame, Eq. (\ref{fast}). At start $h$ is the same for both $m_1$ and $m_2$. Both bodies will have a higher  acceleration than in the inertial case, and also higher than if they had been alone, but their fall will soon differ as the self-force is proportional to either $m_1/M$ or $m_2/M$. This will separate the bodies and prompt a three-body system. All in all, during fall the two bodies will be subjected to different pullings.}
\end{itemize}              

After all, the physics attributed to Aristot\'el\=es is not totally  wrong, and the one to Galilei not entirely right. This stand   complements Rovelli's analysis \cite{ro14} that shows that the Aristotelian physics is a correct and nonintuitive
approximation of Newtonian physics for motion in fluids as Newton's theory is an approximation of
Einstein's theory. 

It should be mandatory when presenting the free fall to state `{\it Since
the mass of the Earth is enormous when compared to the mass of the free falling particle, it is a good approximation to consider the Earth as an inertial reference system.}' \cite{santossoarestort2010}. In this manner, myths, misconceptions and confusions would not rise, and the ranks of those perceiving Galilean free fall as a principle, and not as an approximation as it should be, would not be reinforced.   
 Further, it would be a simple introduction to gauge dependence in general relativity.  

A final remark on the Newtonian self-force. In \cite{depo04}, the singular contribution is extracted and shown as isotropic. Detweiler and Poisson identify the regular part that gives rise to the self-force, Eqs. (\ref{slow2},\ref{fast}).

\section{Free fall in black holes}

Can free fall be used for a phenomenological introduction to general relativity? Let us broach few concepts from general relativity which beauty allows to use just words.  

\ul{Perception of free fall}. Newtonian and Einsteinian visions are opposed. What is the presence of gravity for one is the absence of gravity for the other: 
`{\it Newton's apple hangs in a tree. The force of gravity is balanced by the force from a
branch, and the apple is at rest. Later, the apple falls and accelerates downward until
it hits the ground. [...]

Einstein's apple, being sentient and hanging in a tree, explains its own
non-geodesic, non free-fall, accelerated motion as being caused by the force it feels
from the branch. When the apple is released by the branch, its subsequent free fall
motion is geodesic and not accelerated. The apple is freed from all forces and does
not accelerate until it hits the ground}' \cite{de11}.

\ul{Geodesic}. 
In general relativity, a trajectory dictated exclusively by the main mass, is called geodesic. It corresponds to the trajectory of an inertial observer in pre-relativistic physics. It is characterised by a null acceleration since, according to the EP, to a falling body we can associate an inertial frame; that is to say, that a falling body doesn't feel gravity.  
Put differently, geodesics, forged by the main mass, are spacetime pathways, {\it i.e} the inertial trajectories which test masses are obliged to follow.

\ul{Black hole}. It is a region of space from which nothing, including light, can escape. 
Around a black hole there is a position of no return, called the event horizon. The horizon corresponds to the gravitational radius 
$r_{\rm g}$ of the black hole: if a mass $M$ were compressed in a sphere of radius $r_{\rm g}$, then its gravity would become so strong that nothing at $r\leq r_{\rm g}$, including light, could escape $M$. The concept of light being trapped in a star was presented in 1783 by Michell \cite{michell1784} in front of the Royal Society audience and later by 
Laplace \cite{laplace1796,laplace1799}\footnote{Preti \cite{preti09} describes the close resemblance between the algebraic formulation of Laplace \cite{laplace1799} and the concept of a black hole.}.
In a nutshell, the XVIII concept in modern language was based on the identity of the kinetic energy of a light particle of mass $m$ and velocity $c$, and the potential energy of a body of mass $M$

\beq
\frac{1}{2} m c^2 = G \frac{M m}{r_{\rm g}}
\eeq 

The radius $r_{\rm g}$ that the body $M$ must have to impede light to escape is 

\beq
r_{\rm g} = \frac{2GM}{c^2}
\eeq 

\ul{Multipole moments}. The gravitational monopole of an object is just the total amount of its mass. The gravitational dipole is a measure of how much that mass is distributed  away from some centre in some direction. It is a vector, since it has to convey not only how much the mass is off-centre but also which way. The quadrupole represents how stretched-out along some axis the mass is.

\ul{Gravitational radiation}. It is the transmission of energy by accelerating masses. The amplitude, polarisation and speed of the waves depend upon the specific theory at hand. In case of general relativity, the waves are produced if the second time derivative of the quadrupolar moment differs from zero. In free fall the quadrupolar moment of the system (large and small masses) patently changes in time. That is why falling masses produce gravitational radiation.   

\ul{Einstein equations}. They consist of ten, reduced from the original sixteen  for symmetry of the indices, second order partial and 
non-linear differential equations represented in tensorial form. For our purposes it is sufficient to write the equations as 

\beq
{\rm Geometry} =  \kappa~{\rm Mass}
\eeq
where the left hand-side term is the spacetime geometry in four dimensions curved by the mass on the right hand-side. The constant $\kappa$ is proportional to the gravitation constant $G$ and to the inverse of the fourth power of the speed of light $c$.   

\ul{Metric}. It is a solution of the Einstein equation and represents a given spacetime. In presence of gravity, the most known metrics  are for spherical massive bodies - possibly spinning, and uniform, isotropic, homogeneous distribution of matter, used in cosmology. There isn't an exact solution for two massive spherical bodies, conversely to Newtonian gravity. The metrics are expressed in four dimensions. The interval $ds^2$ is the four-distance in a given spacetime, and it can be intuitively seen as an extension of the Pythagorean theorem to four dimensions.     

{\subsection{The controversy on the geodesic infall}\label{cgi}}

The first solution of the Einstein equations was proposed in January 1916 by Schwarzschild \cite{sc16}, and independently in May by Droste \cite{dro16a, dro16b}, {\it i.e.} the SD solution. This solution is valid for any spherically symmetric and non-rotating mass in vacuum. If the radius of such mass is $r_{\rm g}$, we have a black hole, term coined in 1967 by Wheeler \cite{wheeler67,wheeler68a,wheeler68b}. 
The SD metric, in spherical coordinates ($ct, r,\theta,\phi $, where $c$ is the speed of light) is 

\beq
ds^2 = - \left(1 - \frac{r_g}{r}\right) c^2 dt^2 + \left(1 - {\ds \frac{r_g}{r}}\right)^{-1} dr^2 + r^2(d\theta^2 + \sin^2\theta d\phi^2)~.
\eeq

Despite the mathematical simplicity, a controversy on a falling mass in SD geometry took place from 1916, before decaying after the '80s \cite{sp11,spri14,ei87}. 
The discussion was largely a reflection of coordinate arbitrariness (and unawareness of its consequences), but the debaters showed sometimes a passionate affection to a coordinate frame they considered more suitable for a `real physical' measurement than other gauges. Further, ill-defined initial conditions at infinity, inaccurate wording (approaching rather than equalling the speed of light), sometimes tortuous reasonings despite the great mathematical simplicity, scarce propension to bibliographic research with consequent claim of new  findings, all contributed to the duration of this debate. 

Four types of measurements can be envisaged: local measurement of time $dT$, non-local measurement of time $dt$, local measurement of length $dR$, non-local measurement of length $dr$. Locality is somewhat a loose definition, but it hints at those measurements by rules and clocks affected by gravity (of the SD black hole) and noted by capital letters $T,R$, while non-locality hints at measurements by rules and clocks not affected by gravity (of the SD black hole) and noted by small letters $t,r$. This definition is 
not faultless (there is no shield to gravity), but it is the most suitable to describe the debate, following Cavalleri and Spinelli \cite{casp73, casp77, casp78, sp89}, and thus Sexl \cite{se67} and Thirring \cite{th61}.   
Therefore, for determining (velocities and) accelerations, four possible combinations do exist: 
\vskip6pt
\begin{itemize}
\item{Unrenormalised acceleration $d^2r/dt^2$;} 
\item{Semi-renormalised acceleration $d^2R/dt^2$;} 
\item{Renormalised acceleration $d^2R/dT^2$;} 
\item{Semi-renormalised acceleration $d^2r/dT^2$.}
\end{itemize}
The latter is zero by definition of geodesic, and the discussion will be limited to the first three types. The former two 
present repulsion at different conditions, while the third one never presents repulsion. According to the types of measurement, namely affected by gravity (an observer close to the falling mass) or not (an observer far from the black hole), scholars denied or claimed the existence  of gravitational repulsion.  For a far observer, the most known representation of repulsion is given by an object coming from far away and directed to the black hole but never entering it; less known are the decelerations occurring to particles or photons at given distances from the horizon, or at given velocities. This happens whenever coordinate time $t$ is used, as opposed to proper time $T$. 

The significance of the semi-renormalised quantities might be intriguing. The quantity $d^2R/dt^2$ is measured by a far observer who uses his own clock for time measurements. For distances, the observer uses a meter stick placed at the particle, or compares
the position of objects which are local to the particle, or integrates the echo times of signals reflected by the particle \cite{jash72, jash73}.

The first to introduce the idea of gravitational repulsion was Droste \cite{dro16a, dro16b} himself. He defines 

\beq
dR = \frac{dr}{\sqrt{1-\ds\frac{2GM}{rc^2}}}~,
\label{deltadroste}
\eeq
which, after integration, Droste called the distance $\delta$ from the horizon. 
For radial trajectories, through the Lagrangian and Eq. (\ref{deltadroste}), Droste derives that the semi-renormalised velocity and acceleration \cite{dro16b}

\beq
\frac{dR}{dt} = - c \sqrt{\left(1 - \frac{r_g}{r}\right ) \left[ 1 - A\left(1 - \frac{r_g}{r}\right)\right]}~,
\label{eq:dRdtdroste}
\eeq

\beq
\frac{d^2R}{dt^2} = - \frac{r_g}{2r^2} 
\left[
\sqrt{1- \frac {r_g}{r}} - 
{\ds 
\frac 
{2\left(\ds dR/cdt\right)^2}{\sqrt{1- \ds \frac {r_g}{r}}}} 
\right]= - \frac{r_g}{2r^2} \left[ - 1¨+ 2A\left(1 - \frac{r_g}{r}\right)\right]\sqrt{1- \frac {r_g}{r}}~, 
\label{eq:d2Rdt2droste}
\eeq
where $A$ is a constant of motion ($A = 1$ for a particle falling with zero velocity from infinity)

\[
A = \frac{1}{1- \ds \frac {r_g}{r}} - \frac{(dr/cdt)^2}{\left (1- \ds \frac {r_g}{r} \right )^2}.
\]

$A$ is proportional to $E$, constant of motion representing energy per unit mass ($E = c^2$ for a particle falling with zero velocity from infinity)

\[
E = c \left( 1 - \frac{r_g}{r}\right ) \frac{dt}{dT} = c^2 
\sqrt{1- \ds \frac {r_g}{r_0}}~,
\]

being $r_0$ the coordinate of the starting point.
From Eq. (\ref{eq:d2Rdt2droste}), two conditions may be derived for the semi-renormalised acceleration, for either of which the repulsion (the acceleration is positive) occurs: for $A=1$, if 
${\ds \frac{dR}{dt}} > {\ds \frac{c}{\sqrt{2}}}\sqrt{1- {\ds \frac{r_g}{r}}}  $ or else $ r<2 r_g$. 

Instead in his thesis \cite{dro16a}, Droste investigated the unrenormalised velocity and acceleration and for zero velocity at infinity, they are

\beq
\frac{d r}{dt} = 
- c \left(1 - \frac{r_g}{r}\right )\sqrt{ \left[ 1 - A\left(1 - \frac{r_g}{r}\right)\right]}~, 
\eeq

\beq
\frac{d^2 r}{dt^2} = 
- \frac{r_g}{2r^2} 
\left(1- \frac {r_g}{r} - 3 \frac {(dr/cdt)^2}{1- \ds \frac {r_g}{r}} \right)
= 
- \frac{r_g}{r^2} 
\left(1- \frac {r_g}{r}\right) 
\left[ - 1 + \frac{3}{2}A\left (1- \frac {r_g}{r}\right)\right]~, 
\eeq
for which repulsion occurs if, still for $A=1$,  ${\ds \frac{dr}{dt}} > {\ds \frac{c}{\sqrt{3}}}\left(1 - {\ds \frac{r_g}{r}} \right)$ or else $ r<3r_g $. This coordinate frame is popular, and abundantly used to describe the physics by a far observer.  

The impact of the choice of coordinates on generating repulsion was not well understood in the early days of general relativity. Many notable authors as Hilbert \cite{hi17,hi24}, Page \cite{page1920}, Eddington \cite{ed20}, von Laue \cite {vl21}, Bauer \cite{ba22}, de Jans \cite{dj23, dj24a, dj24b} shared the same convictions and arrived independently to the same formulation of repulsion 
in semi-renormalised or unrenormalised coordinates. Incidentally, they all carried their work ignoring Droste's findings, and often each other. 
Still in the 50s, McVittie \cite{mv56} reaffirms that the particle is pushed away by the central body; likewise do M{\o}ller \cite{mo60}   and Treder \cite{tr72,trfr75}; on the same track Carmeli \cite{ca72,ca82} and others \cite{arifov1980,arifov1981, mcgruder1982}. 
Radar and Doppler measurements with semi-renormalised measurements are proposed by Jaffe and Shapiro  \cite{jash72, jash73}.   

The initial conditions may astray the particle from being attracted by the gravitating mass. Indeed, Droste \cite{dro16b} and Page \cite{pa20} refer to particles having velocities at infinity equal or larger of $c/\sqrt 2$ for the semi-renormalised coordinates and 
equal or larger of $c/\sqrt 3$ for the unrenormalised coordinates. These conditions suggest to Droste and Page that the particle is constantly slowed down when approaching the black hole and therefore impose to gravitation an endless repulsive action. 

Later, von Laue \cite{vl26} writes the radial geodesic in proper time, but it is only in 1936 that Drumaux    
\cite{drumaux1936} fully exploits it. Let us ponder for a while that a geodesic in proper time took twenty years to be proposed! 
Drumaux criticises the use of the semi-renormalised velocity and 
considers Eq. (\ref{deltadroste}) for defining a really physical measurement of length $dR$. Similarly, the relation between coordinate and proper times (for $dr = 0$) provides the physical measurement of time $dT$

\beq
dT = \sqrt{1-\frac{r_g}{r}} dt~. 
\label{drdTdt}
\eeq
Thereby, Drumaux derives the renormalised velocity and acceleration in proper time ($A=1$)

\beq
\frac{d R}{dT} = c\sqrt{\frac {r_g}{r}}~,
\eeq

\beq
\frac{d^2R}{dT^2} = - \frac{GM}{r^2}\sqrt{1 - \frac {r_g}{r}}~, 
\eeq
for which no repulsion occurs. This approach is followed by von Rabe \cite{vr47}, Whittaker \cite{whittaker1953}, Srnivasa Rao \cite{sr66}, Zel'dovich and Novikov \cite{zeno67}, Markley \cite{markley1973}, and others refusing {\it in toto} repulsion. 
We remark instead that even if repulsion is not a gauge invariant phenomenon, it is an interesting feature of the SD metric. An historically oriented account is given by Eisenstaedt \cite{ei87}.  

In conclusion, it was only gauge and there has never been a problem, but we some can't avoid noticing that the controversy decayed only  in the 80s, and that colleagues still address, e.g., repulsion as new concept in physics journals and conferences   \cite{loma09,kuza10,blokvy01,blokvy03}.

\subsection{The research frontiers of free fall in relativistic astrophysics and theoretical physics}
 
\subsubsection{Self-force}

In the '70s, studies on gravitational emission and spectrum began \cite{ze70c}, see the references in \cite{sp11}, but the relativistic self-force in radial fall has been settled only recently \cite{spri14,riaospco2016b}. 
Although foreign to full general relativity, a gravitating point particle is acceptable in the linearised theory. But unavoidably, divergences
will be associated to the infinitesimal size. After all, the difficulties arising from the
point-like description of the mass are to be traded against the simplifications obtained by neglecting
the internal structure.

The mass of the small body itself and the emission of gravitational radiation cause the departure from the geodesic path due to the  self-force, the MiSaTaQuWa equation \cite{misata97,quwa97}.
Conversely, another interpretation retains the concept of geodesic. On the footsteps of Dirac's work \cite{dirac1938}, Detweiler and Whiting \cite{dewh03, de11, po11, popove11} have proposed a novel approach to the self-force, the DeWh approach. Bodies follow the geodesics of an ensemble: field and perturbations. The total field is given by sum of the field of the main body and (the regular part of) the perturbations produced by the small mass and the emitted radiation. So doing, we save the concept of geodesic and EP \cite{spriao14,riaospco2016b}, in the same manner we have retained the EP in Galilean free fall when we have included the mass. 

One main result of the MiSaTaQuWa-DeWh equations \cite{misata97,quwa97,dewh03} has been the identification of the regular and singular
perturbation components and their playing and not-playing roles to motion, respectively. 

\begin{figure}[h!]
    \centering
    \includegraphics[width=0.4\linewidth]{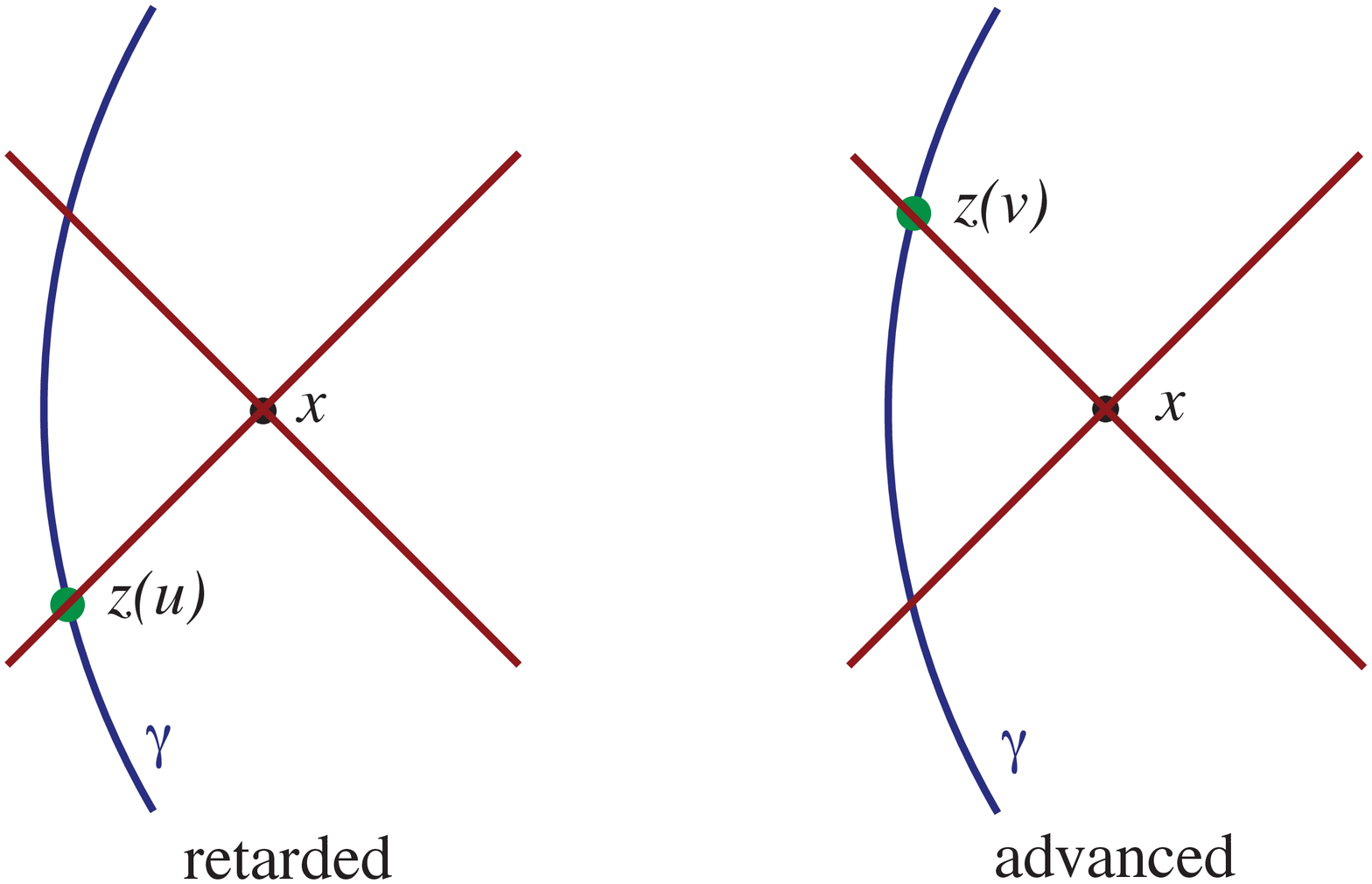}
    \caption{Retarded and advanced terms in flat spacetime.} 
\label{FigLRR}
\end{figure}

We now make a full immersion in Dirac's vision of radiation reaction \cite{dirac1938}, and step from flat to curved space \cite{popove11}. In 
flat spacetime, a negative charge orbits a positive charge Fig. (\ref{FigLRR}). The potential $A^\alpha$ satisfies
the wave equation (FitzGerald-Lorenz gauge condition $\partial_\alpha A^\alpha = 0$): 
$\Box A^\alpha = - 4\pi j^\alpha$.

The $A^\alpha_{\rm ret}$ is the retarded inhomogeneous solution. The radiation goes outward while the charge spirals inward. It 
depends on the state of motion at $z(u)$ (generated by the intersection of the
world line and $x$ past light cone).

The $A^\alpha_{\rm adv}$ is the advanced inhomogeneous solution. The radiation goes inward while the charge spirals outward. It   
depends on the state of motion at the advanced point $z(v)$ (intersection of the world 
line and $x$ future light cone).

The singular term (from the perturbation field) $Sing $ is the mean of the advanced and retarded terms. It is time-reversal invariant, {\it i.e.} incoming and outgoing energy are equal and opposite.  

In flat spacetime, the radiative term is obtained by subtracting the singular from the retarded term. The latter is also singular, but it isn't time-reversal invariant, and shows that the system is losing energy by radiating outward. The subtraction cancels out the singularity at the particle, without any other consequence. Indeed, the singularity is isotropic and it does not exert any force on the particle (in \cite{dewh03}, the authors make an interesting correspondence with the singular contribution in Newtonian physics). It remains only the radiative term to act upon the particle. Conceptually, it is 
given by 

\[
Rad = Ret - Sing = Ret - \frac{1}{2}[Ret + Adv] = \frac{1}{2}[Ret - Adv]~~.
\]      

\begin{figure}[h!]
\centering
\begin{minipage}[c]{0.55\textwidth}
  \includegraphics[width=6cm]{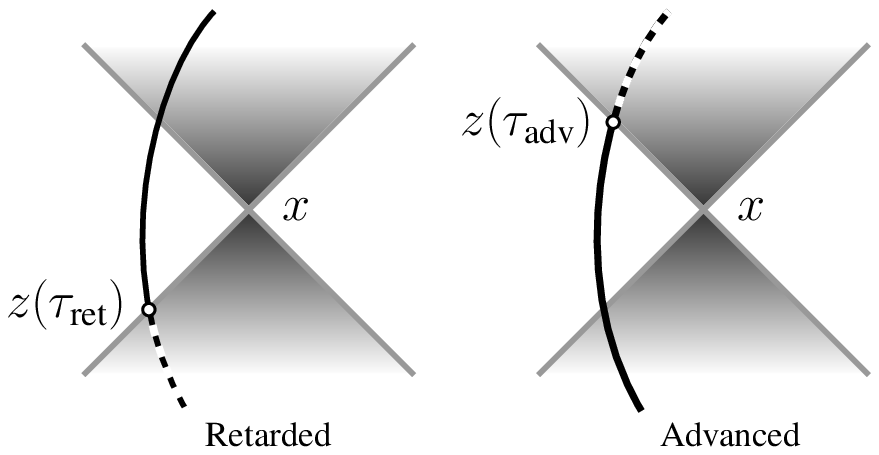}
  \caption{Retarded and advanced terms (dotted line). }
  \label{FigAdvRet}
\end{minipage}%
\quad
\begin{minipage}[c]{0.55\textwidth}
  \includegraphics[width=2.6cm]{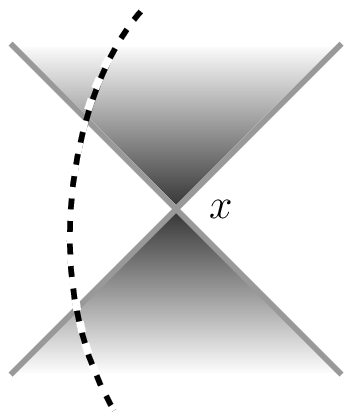}
  \caption{The H term (dotted line). }
  \label{FigH}
\end{minipage}%

\end{figure}

\begin{figure}[h!]
  \centering
\begin{minipage}[c]{0.45\textwidth}
  \includegraphics[width=3cm]{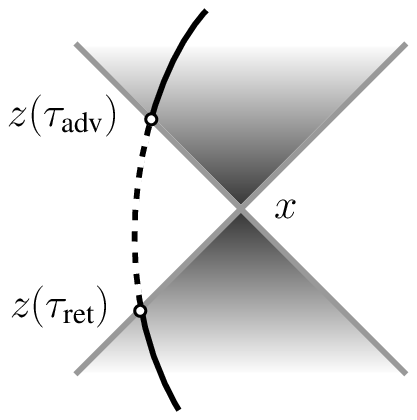}
  \caption{Singular term (dotted line).}
  \label{FigSing}
\end{minipage}%
\quad
\begin{minipage}[c]{0.45\textwidth}
    \includegraphics[width=3cm]{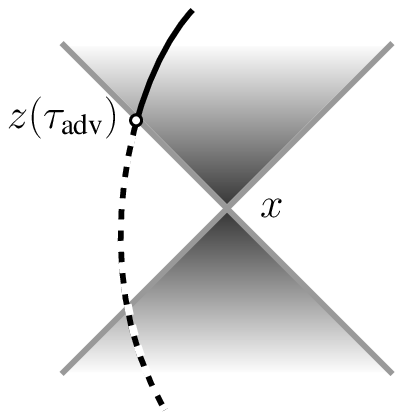}
  \caption{Radiative term (dotted line).}
  \label{FigRad}
\end{minipage}

\end{figure}

We now pass to curved spacetime, Figs. (\ref{FigAdvRet}-\ref{FigRad}). At a given point $x$ the retarded term depends upon the particle's history before the retarded (proper) time $\tau_{\rm ret}$; the advanced term depends upon the particle's history after the advanced (proper) time $\tau_{\rm adv}$.

The singular term depends then upon the particle's history during the
interval $\tau_{\rm ret}<\tau<\tau_{\rm adv}$. 

The straight transposition of the subtraction $Ret - Sing$ to curved space determines still a singularity-free quantity, but the latter  depends upon the contributions from inside of the light cone, past and future. 
The dependence on the future is patently non-causal. The circumvention of this riddle passes through the inclusion of an additional, purposely built, function $H$   

\beq 
Rad =  Ret - Sing = Ret - \frac{1}{2}[Ret + Adv - H] =  \frac{1}{2}[Ret - Adv + H] ~~,
\eeq      
where the {\it ad hoc} function $H$ is defined to agree with the advanced term when the particle position is in the future of the evaluation point, thereby cancelling the $Sing$ term   (the $Ret$ term is zero, for $\tau>\tau_{\rm adv}$). Finally, we have 

\[
Rad_{~\tau>\tau_{\rm adv}} = 0~~.
\]
 
Instead, $H$ is defined to agree with the retarded term when the particle position is in the past of the evaluation point, also cancelling the $Sing$ term  (the $Adv$ term is zero, for $\tau<\tau_{\rm ret}$). Finally, we have 
\[
Rad_{~\tau <\tau_{\rm ret}} = Ret ~~.
\] 

Further, $H$ differs from zero at the intermediate values of the world-line outside the light-cone, between $\tau_{\rm ret}$ and 
$\tau_{\rm adv}$. Thus, the radiative component includes the state of motion at all times prior to the advanced time, and it is not a representation of the physical field but rather of an effective field. Nevertheless, $H$ goes to zero when the evaluation point coincides with the particle position. Figures (\ref{FigAdvRet}-\ref{FigRad}), show the various terms ($x$ the evaluation point, and $z$ the particle position). 

The perturbation $h_{\mu\nu}$ is the difference between 
the full metric of the perturbed spacetime (p)
and the background (b), which, in our case, is the SD metric.
The DeWh approach emphasises that the motion is a geodesic of the metric $g_{\mu\nu} + h_{\mu\nu}^{\rm R} $ where 
$g_{\mu\nu}$ represents the SD background metric, $h_{\mu\nu}^{\rm R} $ the radiative or regular part of the perturbation $h_{\mu\nu}$, and their relation is given in \cite{popove11}. 
It implies two notable features: the regularity of the radiative field and the avoidance of any non-causal behaviour. 

The first order deviation from geodesic motion was determined by Gralla and Wald \cite{grwa08, grwa11}, but see also less technical approaches \cite{spao04,sp11,spriao14}. The coordinate of the particle is $\hat{z}^\alpha = z^\alpha + \Delta z^\alpha$, 
where $\hat{z}^\alpha$ the coordinate of the perturbed trajectory, $z^\alpha$ the coordinate of the unperturbed trajectory, and $\Delta z^\alpha$ the variation. 

The geodesic motion in the b metric is determined by
 
\beq
\frac{D^2z^\alpha}{d\tau^2}= \frac{d^2 z^\alpha}{d\tau ^2} + ^{\rm b}\!\Gamma^\alpha_{\mu\nu} u^\mu u^\nu = 
\frac{d^2 z^\alpha}{d\tau ^2}+ ^{\rm b}\!\Gamma^\alpha_{\mu\nu} \frac{dz^\mu}{d\tau}\frac{dz^\nu}{d\tau}= 0~. 
\eeq   

Through the Dirac-Detweiler-Whiting formulation, the particle crosses the p metric $g_{\mu\nu} + h_{\mu\nu}^{\rm R} $ under geodesic motion     

\beq
\frac{D^2\hat{z}^\alpha}{d\lambda^2}= 
\frac{d^2\hat{z}^\alpha}{d\lambda ^2} + ^{\rm p}\!\!\Gamma^\alpha_{\mu\nu} \hat{u}^\mu \hat{u}^\nu = 
\frac{d^2 \hat{z}^\alpha}{d\lambda ^2} + ^{\rm p}\!\!\Gamma^\alpha_{\mu\nu} 
\frac{d\hat{z}^\mu}{d\lambda}\frac{d\hat{z}^\nu}{d\lambda} = 0~.
\eeq
                                      
The proper times $\tau$ and $\lambda$ refer to b and p metrics, respectively.
Computing the difference between these two geodesics, we get the second order proper time covariant derivative      
\cite{spriao14} 

\beq
m \frac{D^2 \Delta z^\alpha}{d\tau^2} =  ~~~~ - 
\underbrace{ {m~~R_{\mu\beta\nu}}^\alpha u^\mu \Delta z^\beta u^\nu}_{\rm Background~geodesic~deviation~BGD} 
- ~~~~\underbrace{ \frac{m}{2}
(g^{\alpha\beta} + u^\alpha u^\beta) 
(2h_{\mu\beta ;\nu}^R- h_{\mu\nu ;\beta}^R) u^\mu u^\nu}_{{\rm Self-force~SF}~=~F^\alpha_{\rm self}}~, 
\label{gweq}
\eeq
where $R_{\mu\beta\nu}^\alpha$ is the Riemann tensor, and $u^\alpha$ the four-velocity. In Eq. (\ref{gweq}), the BGD term does not contain perturbations, and depends solely on the SD metric, while the SF term      
represents the self-force. In a sense, the SF term produces the BGD term which wouldn't exist in absence of the SF term. Then, we would have 
~$
m {\ds \frac{ D^2 \Delta z^\alpha}{d\tau^2}} = 0, 
$
~meaning that there is no displacement, since we have neglected the small body own mass and emitted radiation. 

To get meaningful self-force quantities, the regularisation of the divergencies is mandatory. It is an highly technical problem for which we refer to the Mode-Sum method proposed by Barack and Ori \cite{baor00,baor01}.

One of the main concerns for gravitational wave detection is the capability to simulate waveforms with great accuracy. Gralla and Wald consider that any perturbation scheme is doomed to failure at
late times due to the increasingly growing BGD term. Thus, they suggest to evolve the most relativistic orbits through the iterative
application of the SF term on the particle world-line, {\it i.e.} the self-consistent
approach \cite{grwa08,grwa11}. In radial fall, the self-consistent method was applied \cite{spri14,riaospco2016b}. 

The conclusions on the fully relativistic radial fall of a small body towards a large one can be summarised as follows. 
In specific, but popular coordinate systems in general relativity, {\it i.e.} Regge-Wheeler, harmonic and all others smoothly related gauges, a far observer would conclude that the self-force pushes inward (not outward) the falling body, with a strength proportional to the mass of the small body for a
given large mass, that is $m/M$. Thus the heavier body will reach the SD black hole horizon premises before the lighter companion. 

It would be of interest to study the fall in other coordinates. It is plausible to envisage that in other gauges, the self-force would be repulsive, rather than attractive as it occurs in the Regge-Wheeler and harmonic gauges.      

\subsubsection{Black hole evaporation}

The difficult combination of general relativity and quantum mechanics in black holes  provides us of insights in research frontiers \cite{harlow2014}. Indeed, despite being firmly classical objects, black holes offer a glimpse into the nature of quantum gravity. One of the reasons is that, in general relativity, black holes are described by a metric which is regular at the horizon, but singular inside. Therein, Einstein's theory ceases to be valid and quantum corrections are needed. 

Classically, the only measurable quantities are the black hole total mass, spin and electric charge. What about thermodynamics? 

Bekenstein \cite{bekenstein1973} identified entropy with the area of the horizon. This result led Hawking \cite{hawking1975} to compute the associated black body radiation, showing a finite horizon temperature and entropy, confirming the latter as one-fourth of the surface area. 

Black holes spacetimes should be consistent with the laws of thermodynamics. As a hot gas accretes onto a black hole, its entropy is absorbed. The microscopic random properties of the gas would no longer be observed once the black hole had absorbed the gas and settled down to rest. The decrease in entropy in the Universe outside the horizon must be compensated by the increase of the entropy of the black hole that swallowed the object. Thus, the horizon area is a non-decreasing function of time. On the other hand, due to Hawking's radiation, black holes radiate, which causes both the black hole's mass and the area of its horizon to decrease over time.

For a black hole of mass $M$, the Bekenstein-Hawking entropy is 

   \begin{eqnarray}
   S=\frac{1}{4} \frac{k A_H}{l_p^{2}}~,
   \label{EQN_S}
   \end{eqnarray}
   where $k$ is the Boltzmann constant, $A_H=4\pi r_g^2$ the horizon surface area, while the Planck area is  $l_p^{2}=G\hbar/c^3$, $\hbar$ being the reduced Planck constant; in words, the entropy of a black hole is one quarter its horizon area in Boltzmann-Planck units.

The horizon prevents an external observer from receiving detailed information on the black hole: how and when it formed in some prior epoch of the gravitational collapse.  Meanwhile, black hole entropy is a measure of the missing information about the past. Shannon entropy \cite{shannon1948} is the expected value (average) of the information contained in each message.  

In quantum mechanics, the complete information about a system is encoded in its wavefunction. The evolution of the wavefunction is determined by a unitary operator, and unitarity guarantees that information is maintained in the quantum sense. Therefore, viewed as classical objects, black hole formation woefully violates unitarity by the appearance of amounts of entropy. To save unitarity, the information quantified by entropy should somehow be recoverable. 

   Equation (\ref{EQN_S}) has suggested to 't Hooft and Susskind \cite{thooft1993,susskind1995} that black holes are objects emerging in a thermodynamic limit of a microphysical representation of spacetime. If so, spacetime and matter effectively emerge holographically from information on two-dimensional screens. Accordingly, the spherical screen of area $A=4\pi r^2$ requires
   a Shannon information $I$ \cite{vanputten2015a}

   \begin{eqnarray}
   I = 2\pi \Delta \phi~,
   \label{EQN_I}
   \end{eqnarray}
   defined by the phase difference $\Delta\phi=r_gr/2l_p^2$ in wavefronts of de Broglie's wavefunction of $M$.
 
For $A_H=4Il_p^2/k$, Eqs. (\ref{EQN_S}) and (\ref{EQN_I}) get equal. As $I$ is now classically hidden from view by the no-hair theorem of black holes in general relativity \cite{carter1971,hawking1972,hawking2005}, Eq. (\ref{EQN_I}) becomes the entropy of Eq. (\ref{EQN_S}).
   
As Hawking argued \cite{hawking1974}, if black holes have a finite entropy, they have a finite temperature, by which they may come to 
equilibrium with a thermal gas of photons. Black holes would hereby absorb and emit photons alike, and they
would emit them in the right amount to keep balance. Hawking radiation is the most striking demonstration of quantum mechanics at work in general relativity. Hawking radiation displays as particle-antiparticle
pair appearing close to the horizon of a black hole. One of the pair falls into the black hole while the other escapes. For preserving total energy, the particle that fell into the black hole must have had a negative energy (with respect to a far observer). By this process, the black hole loses mass, and, to an outside observer, it would appear that the black hole has just emitted a particle.  


   Hawking radiation is qualitatively similar to the evaporation of globular clusters of stars, see \cite{vanputten2012a} and references therein. Self-gravitating systems have negative heat (the more mass and energy a system absorbs, the colder it becomes. In contrast, if it is a net emitter of energy, it will become hotter and hotter until it boils away). To leading order, the velocity distribution of their constituents evolve to a Boltzmann distribution, whose tail inevitably contains escapers with positive total
   energy. Quantum mechanics permits the escape of radiation from black holes, 
   like high velocity stars escape from globular clusters. By negative specific heat, evaporation gradually 
   accelerates, leaving a finite lifetime of black holes and globular clusters alike. 
  
   In practical terms, the lifetime of macroscopic black holes is astronomically large, as 
   Hawking radiation proceeds by exceedingly slow photon emission one-by-one every few thousand light 
   crossing time scales $t_c=r_g/2c$ \cite{vanputten2015b}. A one solar mass black hole hereby assumes a 
   photon emission rate of about $40$ Hz.

   The exceedingly slow emission rate and the implied long evaporative lifetime is a 
  direct consequence of unitarity. For unitarity to be saved, there must be a one-to-one relation between the final remnant in Hawking radiation, after complete evaporation of the black hole in the distant future, and the initial conditions 
   leading to collapse to a black hole in the distant past. 


%

  Distant observers see the falling object coasting on the event horizon, effectively smeared out in its mass (and angular momentum 
  and electric charge) uniformly over the event horizon, according to the no-hair theorem of black holes in general
  relativity. For these observers, evaporation is so slow, that it can be safely neglected.
  In contrast, the object sees things very differently. Submerged at deep redshift and relativistically moving
  inwards as seen by local static observers hovering close by, it experiences no event horizon
  by Einstein's EP \cite{wald1984}. 

In \cite{vanputten2015b}, we propose that principle back reaction on
the evolution of the black hole spacetime occurs in discrete steps according to 
   \begin{eqnarray}
    \dot{M} \propto - \sum_{i=1}^n \epsilon_i \delta(t-t_i)~, 
    \label{EQN_M}
  \end{eqnarray}
  where the $\epsilon_i$ and $t_i$ refer to the energies and rare instances of photon emissions observed at asymptotic infinity. 
  By virtue of the low rate of emission,  Eq. (\ref{EQN_M}) is an effective description for an extended region of
  SD spacetime around the black hole. 
The total energy at infinity $E$ of the proverbial apple in free fall onto a black hole will no longer be constant but, rather, be experiencing changes similar to those in $M$. 
  

Falling massive observer trajectories are described by geodesics containing specific jump conditions in the velocity four-vector at instances of photon emission, Eq. (\ref{EQN_M}). A detailed calculation \cite{vanputten2015b} proposed a new conserved quantity in the form of the {\em product} 
  \begin{eqnarray}
  EM=\mbox{const.}~,
  \label{EQN_EM}
  \end{eqnarray}
  valid for $k_BT_H < E\ll M$. It shows that the total energy of the falling object, a relativistic and quantum apple, is increasing in time for a far observer.  
%

What about the local observer sitting on the apple? He will be harder and harder to define, as its local gravitational acceleration diverges at photon emission, whence is subject to a divergent temperature \cite{fulling1973,davies1975,daviesfullingunruh1976,unruh1976}.

   Free fall is a powerful probe for evaporating black holes and 
general relativity no longer clashes with unitarity (and hence with quantum mechanics).
The quantum observer falling in, like its classical colleague, does not feel as if it enters an event horizon. In fact, the observer never quite reaches the evaporating horizon. Thereby, he agrees with the observer at infinity, the black hole disappearing under his eyes, for the Hawking radiation.  

\section{The relativistic Pisa tower}

Let us return to Earth and treat the fall relativistically. If benefitting of the ideal technology for gravitational radiation detectors, could we observe gravitational waves emitted by bodies falling from the height of the Pisa tower to the ground? A sort of relativistic version of the legendary experiment? The upper limit for the energy radiated would be approximately given by \cite{ze70c}

\beq
E_{\rm gw} \sim 10^{-2}m c^2 \frac{m}{M}\left(\frac{r_{\rm g}}{r_{\oplus}}\right )^{\frac{7}{2}}\approx 10^{-41}~{\rm J}~.
\eeq

For a fall from the Pisa tower, the product $\Delta E \Delta t$ turns to be orders of magnitude below $\hbar/2$ (where $\hbar$ is the reduced Planck constant), and thus beyond the measurability threshold dictated by the indetermination principle of Heisenberg. Needless to add, the different falling times due to the Einsteinian self-force would be of higher order with respect to the differences at Newtonian level. The reasons of non-detection are though of profoundly different nature: in the Newtonian case, the measurement is impossible for the state of the art of technology, in the Einsteinian case for the foundations of quantum mechanics.

\section{Conclusions}

At the centennial of general relativity, we have examined the free fall. Exception made for the inertial frame, in pre-relativistic physics and respecting the equivalence principle, bodies fall slower or faster according to their mass and the chosen coordinate system. 
Thereby, we have introduced the concept of gauge, as a fore-runner of the general relativistic problem.

Indeed, different coordinates induce different observers to draw different conclusions from their experiments. The unawareness of the role of the coordinates played a large role in the debate that took place mainly up to the early 80s concerning the free fall into a black hole. 

We have shown that equivalence principle can be forged such that the inclusion of the mass of the falling body does not invalidate it, neither in Galilean nor in the fully general relativistic free fall. In the latter, bodies keep on falling with an acceleration dependent on the amount of mass and gravitational radiation emitted. Nevertheless, free fall can be described by a geodesic equation, even when the mass of the falling body is considered. 

In analysing free fall into evaporating black holes, we have argued on a constant of motion compatible to general relativity and quantum mechanics descriptions.   

We have also imagined that ordinary masses are dropped from a Pisa-like tower, and evaluated Galilean and Einsteinian falls.
The state of the art of technology is not yet capable to measure the difference in the fall of ordinary masses. Instead, though general relativity predicts the generation of gravitational radiation even in a $60$ m stretch, it is remarked that the Heisenberg  uncertainty rule, though, impedes the detection of any such radiation.  

Finally, science and technology, could be portrayed in their historical and epistemological development, by referring to the progressive understanding of free fall throughout the centuries. 
The epistemological significance, as the legendary aspects related to Aristot\'el\=es, Galilei, Newton, and Einstein, all major figures who have dealt with radial fall, would of be interest to a larger public. 
The support by animation devices would enhance the message and help reaching unskilled audiences.     

\section{Acknwoledgements} 
MHPMvP acknowledges stimulating discussions with J. Polchinski and G.'t Hooft; ADAMS with S. Gralla and R. Wald.

{\section{Appendix}\label{appendix}}

\subsection{Misconceptions on Galilean free fall}

In the literature, the contribution by air friction is by far the major concern which may provide a result different from equal falling times.  
The difference between fall in vacuum and in the air has been the subject of a polemics between the former French Minister of Higher Education and Research Claude All\`egre and the Physics Nobel Prize Georges Charpak, solicited by the satirical weekly `Le Canard Encha\^in\'e' \cite{canard99}. The Minister affirmed on French television in 1999
`{\it Pick a student, ask him a simple question in physics: take a petanque and a tennis ball, release them; which one arrives first? The student would tell you: ``the petanque''. Hey no, they arrive together; and it is a fundamental problem, for which 2000 years were necessary to understand it. These are the basis that everyone should know.}' The humourists wisecracked that the presence of air would indeed prove the student being right and tested their claim by means of filled and empty plastic water bottles being released from the second floor of their editorial offices ...and asked the Nobel winner to compute the difference due to the air, whose influence was denied by the Minister. But in this polemics, no one drew the attention to the Newtonian self-force, also during the polemics revamped in 2003 by All\`egre \cite{allegre03} who compared this time a heavy object to a paper ball. 

An approximation is a valid shortcut as long as it is perceived as such. When an approximation is presented or perceived as principle, misconceptions arise. High impact science, physics education journals and media not always state that the uniqueness of free fall is solely valid in an ideal inertial frame. 


In the physics education literature, out of the many references examined, we have found only very few exceptions. Lehavi and Galili state
`{\it However, the requirement that the falling body be relatively small is rarely stated in an educational context.  Typically, Galileo's law is quoted just as he stated it  (as an empirically proved conjecture) [...]. These} [advanced mechanics] {\it courses treat the two-body problem, which includes the context of "falling," but do not revisit Galileo's law. Even a discussion of free fall in The Physics Teacher (1996, 37) ignored the possibility of its mass dependency.}' \cite{lega09}; and in Phys. Teach., de la Vega bluntly affirms `{\it We read, in most physics books, that all objects fall toward the center of the earth with the same acceleration, no matter what their masses are. I believe that this is wrong.}' \cite{delavega1978}. 
Indeed `{\it Galileo's law of free fall is always presented as the appropriate, unlimited claim
against which the misconception that heavier objects fall faster than lighter ones is measured'} \cite{lehavigalili2008}, and 
` {\it ignoring the approximate nature of Galileo's law in the modern teaching of physics presents a
conceptually deficient instruction}' \cite{galili2009}.

This lack of attention is somewhat surprising as it appears that education programmes are often unsatisfactory, in this respect. In \cite{lega09}, we read that the `{\it evidence of several areas of confusion regarding Galileo's law of free fall. This confusion indicates short comings in the instruction of introductory physics courses. The uniformity of the answers we received indicates that Galileo's law is understood as a general principle and not as an approximation. This conception is extremely strong, and our attempts to stimulate its refinement during the interviews failed. [...] Students thought that it was resolved in advanced physics theories, but teachers were confused and expressed much concern about their level of expertise. As one of the interviewees said: `I cannot argue against the whole community' Our interviewees preferred to recognize their own failure, assuming that they just missed `the point'.}'.

Lehavi and Galili  analyse the responses of students and teachers confronted with a large questionnaire concerning several issues in classical mechanics; they 
recognise the problem of the Galilean free fall be perceived as a principle, but they do not address mass dependence sufficiently, nor they suggest a teaching approach. Like de la Vega \cite{delavega1978}, they show a reference frame where mutual acceleration occurs, Eq. (\ref{fast}), but neglect the centre of mass displacement, Eq. (\ref{slow2}). 

We present the results of an analysis of the four major physics education journals (American Journal of Physics,  European Journal of Physics, the Physics Teacher, and Physics Education), together with the responses of a questionnaire filled in by physics and chemistry undergraduates in Orl\'eans.  

Outside the education world, and namely in the research literature, one would expect an unambiguous description of free fall. A thorough analysis of two high impact factor journals, Nature and Science, has been carried out. Likewise for physics education journals, we have searched for keywords in the full text, {\it e.g.}, free fall, equivalence principle, Galileo Galilei, Pisa tower. We have found wrong statements, strongly or slightly deceptive omissions in old and recent contributions, sometimes oriented to historical accounting.  
The authors in these journals are either professional physicists who may consider unnecessary or trivial to pay attention to obvious details, or unaware science journalists who, in their youth, might have been victims of sloppy teaching.    

\subsubsection{A case study}

In successive years an ensemble of 54 third year undergraduate students in Orl\'eans were asked to comment whether some statements on the EP are true or false. The questionnaire was submitted to students attending a course on Special Relativity including a brief introduction to General Relativity, Tab. \ref{tab1}. It is remarkable that the highest number of wrong replies is associated to question 5, which is dealt with in this work.  

\onecolumngrid

\begin{table}[H]
\centering
\caption{Questionnaire on EP submitted to 54 third year (mostly physics majors with few chemistry majors) undergraduate students.}
\vskip 10pt
\label{tab1}
\begin{tabular}{|p{7cm}|c|c|c|}
\hline
{\bf Is the statement true or false?}                                
& {\bf Correct replies (\%)}                        
& {\bf Absence of replies (\%)}                    
& {\bf Wrong replies (\%)}                  
\\ \hline
{\small 1. The gravitational and inertial masses are equivalent, according to the EP.}
& 44                            
& 0
& 56
\\ \hline
{\small 2. The tidal force obliges freely falling bodies to approach. Thus the EP is valid only locally.}
& 78                           
& 0
& 22
\\ \hline
{\small 3. The bending of light grazing a mass is due to the EP.}
& 67
& 11                           
& 22
\\ \hline
{\small 4. A weight of a given material falls with the same acceleration of an equal weight of another material, according to the EP.}
& 67                            
& 11
& 22
\\ \hline
{\footnotesize 5. In any reference system, two masses of the same material but different weights fall with the same acceleration.}
& 33                            
& 0
& 67
\\ \hline
\end{tabular}
\label{orleans}
\end{table}

\subsection{High impact factor and physics education journals}

We pick few examples from high impact factor and from physics education journals. Statements like `{\it [...] the acceleration of a body in the presence of only a gravitational field is independent of its mass (g-universality)}' \cite{ia87}, or `{\it Equipped with a cannonball and a musket-ball, Galileo climbed the steps of Pisa's famous leaning tower, and, together with the intelligentsia of that glamorous Italian city, watched them drop to the ground at the same speed. In one simple stroke he had disproved Aristotle's notion that objects fall at a rate that depends on their bulk}' \cite{ba05} or `{\it [...] this miracle of physics (the physicist who does not delight in seeing a 50-cent piece and a penny fall together
has no soul)}' \cite{fa67} create an unnecessary myth. There is no miracle, but just misrepresentation;  
among the many examples of an apodictic repulsion of mass dependence `{\it [...] the heavier falls faster (farther) in proportion to its weight. Of course, Aristotle's law is not true [...]}' \cite{hallounhestenes1985}. The role of coordinates appears to be unknown. 

We note that even a different world and God are evoked in \cite{wereley1988}. Wereley analyses the logic of Galilei's arguments exposed by Simplicio and Salviati in a {\it reductio ad absurdum} (RAA) form to conclude \'{\it Is Galileo's argument sound? If it is, there is no possible world where heavy objects could fall faster than light ones. Not even God could create such a world if the RAA is sound.}'. But then he withdraws from his original stand and concludes '{\it Galileo's RAA argument is not sound and thus he has not proven that is not possible to have a world where heavy objects fall faster than lighter ones}.   
  
In many cases the statement though scientifically accurate `{\it We show that the macroscopic glass object used in this instrument falls with the same acceleration, to within $7$ parts in $10^9$, as a quantum-mechanical caesium atom}' \cite{pechch99} might reinforce less attentive readers' misconception that they are confronted with a principle of physics, and not just an approximation associated to a specific coordinate frame.   

There are also instances in which a strenuous defense of Aristot\'el\=es (both in high impact factor \cite{ha14} or education literature \cite{franklin1979}) leads almost to ascribe the fall at the same rate of bodies of different masses to the ancient Greek philosopher. 

Many references consist of the research work by historians who were mainly concerned whether or not the Pisa free fall experiment was actually carried out, by who and under which conditions. The efforts in interpreting Galilei's writings have not clarified the issue of mass dependence. In \cite{adlercoulter1978}, Adler and Coulter discuss the role of density of the falling mass. They refer to \cite{drabkindrake1960} to state that Galilei's initial view was such that objects of same size, but different material, would acquire a larger speed in vacuum, conversely to the view held later on \cite{drake1974}.   

But the main question appears unanswered. Is the equality of free fall an exact or an approximate statement? Therefore, one main purpose of this essay is to get the record straight. 

Science educators, communicators and researchers should be attentive to these arguments, {when teaching Galilei's statements on free fall.} 
Approximations are to be avoided, like `{\it The equivalence between inertial mass (in m{\bf a}) and gravitational mass (in m{\bf g}) is a powerful principle with remarkable, counterintuitive consequences.
It makes a hammer and feather fall together in vacuum —- or on the Moon [...]}' \cite{peekhamaoury14}. This occurs although explicit attempts of clarifications were made in the physics education literature just for this unfortunate conclusion  `{\it 
A condition that is commonly made explicit addresses air resistance. Teachers often show footage of an astronaut dropping
a hammer and a feather on the Moon to demonstrate that "Mr. Galileo was right" (Exclaimed by astronaut Dave Scott standing on the Moon in 1971 Apollo 15 mission).}' \cite{lega09}. 

Quotation of Galilei's writings should always be accompanied by the argument that equal times in free fall of different masses occur only in an inertial frame. Maybe, Galilei was aware, maybe not, but we definitively should.  


\subsubsection{Analysis of the journals in physics education}

For the analysis of physics education literature on Galilean and generally pre-relativistic classic free fall, we have disregarded those papers exclusively addressing historical topics or experimental measurements, in air or vacuum. Instead, we have evaluated all those papers where there has been sufficient discussion on the more scientific aspects of the free fall. Further, free fall is mentioned countlessly in papers in subjects other than classical mechanics, or even other than physics (incidentally, most of the time, free fall is raised to a mythical state therein). We have also disregarded the works in these latter two domains. Finally, there is no claim of having spotted all references. We have limited our search to the following journals: 
American Journal of Physics - formerly the American Physics Teacher (AJP) - from 1933, European Journal of Physics (EJP) from 1980, the Physics Teacher (PT) from 1963, and Physics Education (PE° from 1966, and defined the following categories:
\begin{enumerate}
  \item {Category I. Papers addressing mass dependence in the usual terms of mutual acceleration, Eq. (\ref{fast}), but omitting Eq. (\ref{slow2}). Thereby, the issues of coordinate freedom and absence of gauge invariance are ignored \cite{delavega1978,mallmannetal1994,lega09}. }
  \item {Category II. Papers omitting to state that acceleration mass independence occurs solely in the inertial frame \cite{dellavalle1974,adlercoulter1978,bondi1986,wereley1988,gallantcarlson1999,gallant1999,french1999,mallinckrodt1999,editorpt1999,borghietal2005,hewitt2005,brown2006,foong2008,chtepeugja14,khprako2015}.}
  \item {Category III. Papers omitting to state that acceleration mass independence occurs solely in the inertial frame and manifestly arguing that mass independence derives from the EP \cite{pockman1951,burnistonbrown1960,burnistonbrown1976,egdall2009,baku11,nolucrshtupecaanza13,peekhamaoury14}.}
  \item {Category IV. Papers affirming (implicitly or explicitly) that mass dependence is (always) a manifestation of an erroneous view \cite{lindsay1942,cunninghamkarplus1962,feinberg1965,seeger1965,raman1972,casper1977,gee1977,redding1978,franklin1976,adlercoulter1979,franklin1979,champagneetal1980,nachtingall1980,whitaker1983,lythcott1984,hallounhestenes1985,ryder1987,kleinmittelstaedt1997,disygarner1999,hsu2001,trainer2005,art2006,coronaetal2006,pantanotalas2010,perssonhagen2011,verarivera2011,kozniak2012,abdelazemal-basheer2015,khprako2015}.}      
\end{enumerate}

The assignment of a given work to one or another category is sometimes dubious, especially for categories II and IV, but we have refrained to assign multiple categories to draw a clean statistics. Obviously, there is some arbitrariness in this process. 

In textbooks, the situation is also quite unsatisfactory \cite{lehavigalili2008,lega09} and the conclusions remain similar.  
Finally, we counted 21  
references for Nature from 1869 \cite{ja79,bryan04,du08,gr14,ha14,ra14,gu35,ev36,co36,dcd38,xx39,ed46,ro64,ra75,gr86,ia87,pechch99,or00,ll01,ba05} 
and 24 for Science from 1880
\cite{do02,th02,cl03,ni11,hu16,hu20,ca20a,pa20,ca20b,ho21a,ca21,ho21b,we21,co26,ze28,tu29,la37,dsp59,ha65,fa67,ed77,ho91,wa99,ra11} that could be ranked in categories II, III, IV. 

\subsection{Causes of the state of affairs and remedies}
  
Which are the causes of this state of affairs? 
We estimate that in high school teaching, as well as in the media, the difficulties associated to subtle concepts as centre of mass displacement, mutual acceleration, non-inertial observers, gauge dependence, and the Newtonian self-force induce to oversimplify the presentation of free fall `{\it Why don't we discuss these }[Newtonian gravity self-force] {\it effects in undergraduate classical mechanics? The primary reason is that the Newtonian two-body problem can be solved easily and analytically without mention of the self-force. But in addition, a description of the Newtonian self-force introduces substantial, unavoidable ambiguities which are similar to the relativistic choice of gauge.}' \cite{de11}. These difficulties determine that high school textbooks skip discussing these matters. 
Afterwards, in university courses, the equivalence principle is often presented already in a general relativity context, giving for granted that students have already acquired a critical vision, and thereby not filling the gap between the two education programmes.

There are two dichotomic obstacles. Some physicists immediately grasp the physical concepts and considers them almost elementary (and they are right!). Meanwhile, they tend to be unaware or scarcely receptive of the existence of an educational issue. For them, the problem doesn't exist. 
Others perceive the issue of coordinate dependence in Galilean free fall as complex, awkward  and ultimately wrong. They simply miss the point or are confused by the often oversimplified, and unwillingly brainwashing, literature.     

%
%

We propose that the students and the general public, targeted by the media, be made aware that affirming that lighter and heavier masses fall at the same rate is only a (very) valid  approximation in an idealised frame. It would prevent from nurturing some degree of confusion, or even the risks of cultivating a stereotyped know-it-all posture. Further, the dependence of the fall on the reference frame could be used as introduction to the gauge issue in general relativity, as we have seen in the main body of the text.  

\bibliography{references_spallicci_160709}

\end{document}